# Statistical model of evolution of brain parcellation

Daniel D. Ferrante, Yi Wei, and Alexei A. Koulakov

Cold Spring Harbor Laboratory, Cold Spring Harbor, NY, 11724. USA.

We study the distribution of brain and cortical area sizes [parcellation units (PUs)] obtained for three species: mouse, macaque, and human. We find that the distribution of PU sizes is close to lognormal. We analyze the mathematical model of evolution of brain parcellation based on iterative fragmentation and specialization. In this model, each existing PU has a probability to be split that depends on PU size only. This model shows that the same evolutionary process may have led to brain parcellation in these three species. Our model suggests that region-to-region (macro) connectivity is given by the outer product form. We show that most experimental data on non-vanishing macaque cortex macroconnectivity (62% for area V1) can be explained by the outer product power-law form suggested by our model. We propose a multiplicative Hebbian learning rule for the macroconnectome that could yield the correct scaling of connection strengths between areas. We thus propose a universal evolutionary model that may have contributed to both brain parcellation and mesoscopic level connectivity in mammals.

The brain has many distinct regions defined anatomically and functionally. The evolutionary origin of the diversity of brain regions is not well understood (*1*). According to one theory, new brain regions emerge from existing ones through the process of fragmentation and specialization (*1, 2*). Due to the abundance of brain regions, fragmentation is expected to be an iterative process persisting through brain evolution. One can infer properties of this process from the distribution of resulting fragments, i.e., brain region sizes. Here we examine the distributions of brain region sizes, called here parcellation units (PU), for three species: mouse, macaque, and human. We infer parameters of the brain fragmentation process that can lead to these distributions. We argue that brain fragmentation followed a similar evolutionary mechanism in the three species analyzed.

Interestingly, the problem of brain parcellation is mathematically related to the fragmentation of shells of explosive projectiles and warheads as well as to rock grinding and crushing, analyzed previously in (*3-5*). In explosive shell and rock fragmentation, one original piece gives rise to a distribution of fragments that independently undergo further crushing. The class of processes with independent fragmentation was first analyzed by Andrey Kolmogorov. Kolmogorov showed that under appropriate conditions, sequential breakage yields a lognormal distribution of particle sizes (mathematician was intrigued by the lognormal distribution of gold particles (*6*)). The continuous crushing of rocks and explosive shells, just like the repeated subdividing of brain regions, results in an evolving distribution of cluster sizes, which can yield information about the underlying process. In the present study, we both evaluate the statistical distribution of brain region sizes and propose an evolutionary model that is somewhat distinct from Kolmogorov's theory. We also study the implications of brain evolution by fragmentation for region-to-region connectivity.

PU volume data is available for the three species: mouse (*7*), macaque (*8*), and human (*9*). For the mouse data, it is possible to reconstruct the tree formed by the segmentation of the brain into PUs, as shown in Figure 1. Both branch points and leaves of the parcellation tree reside at different depth (distances to the root), suggesting that the tree is unbalanced (*10*). To analyze the distribution of PU sizes, we fit the distributions of the logarithm of brain region volumes for all three species with Gaussian distributions. Overall, the distributions appear to be close to normal, suggesting that the distribution of PU sizes is close to lognormal (Figure 2). The standard deviations of the logarithm of PU sizes are similar for all three species ($\sigma = 1.47$, 1.24 and 1.40 natural logarithm units for mice, macaques and humans respectively). Statistical tests (Kolmogorov-Smirnov (KS) (*11*)) and quantile-quantile (QQ) plots (Figure 2) show that the distributions of the logarithms of PU size are close to normal ($p_{KS}$=0.07, 0.48, 0.97 for the three species). Larger p-values indicate that the observed distributions of brain region sizes are close to lognormal distributions.

<a2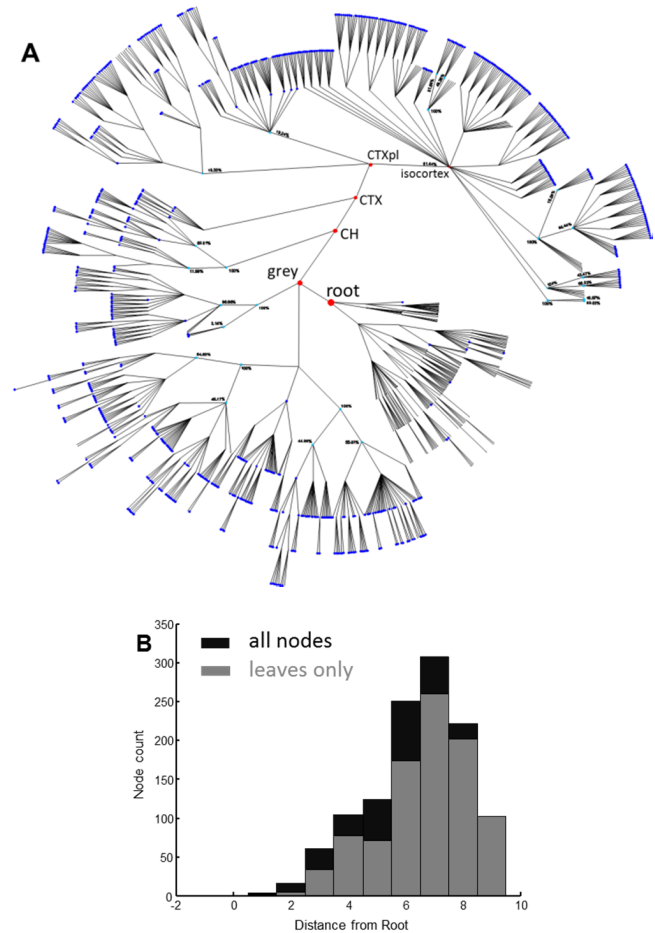

**Figure 1. Mouse brain parcellation tree.** (A) Each brain region (PU) is represented by a node. PUs further from the root are parts of upstream PUs. Leaves of the parcellation tree are included into the histogram in Figure 2A. (B) The mouse parcellation tree is unbalanced. Leaves are abundant at all depths in the tree.

We propose a simple evolutionary parcellation model that can explain these observations. Our process starts with a *tabula rasa* brain containing only a single region (Figure 3A, step 1). This region is then divided into two PUs of equal size (step 2). In the next step, we choose one of these two regions with equal probability and divide it again into two equal parts (step 3). This sequence of steps, including picking a random region independently of its size and dividing it, is repeated until the target number of PUs is achieved (Figure 3). Iterating this model results in a distribution of brain region sizes which is close to log-normal (Figure 3).

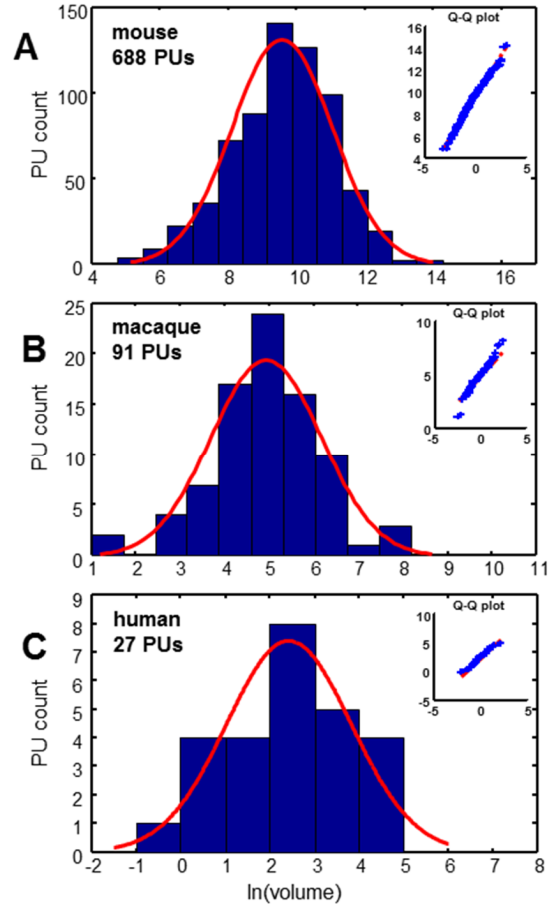

**Figure 2.** Distributions of PU volumes are close to lognormal. Red curves show Gaussian fits. Q-Q plots indicate the high quality of fits. **(A)** Mouse brain: σ=1.47 **(B)** Macaque cortex: σ=1.24 **(C)** Human cortex: σ=1.40.

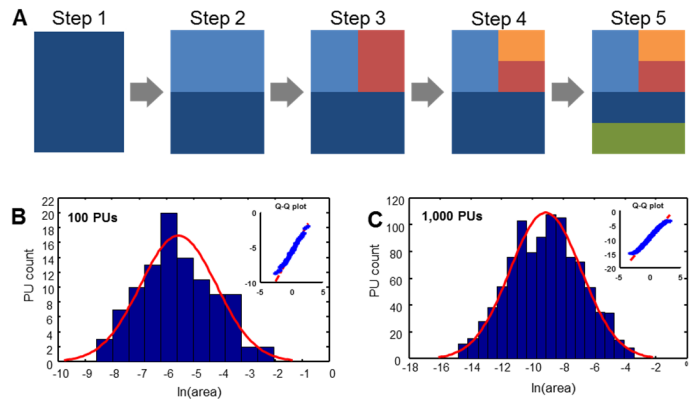

**Figure 3.** Evolutionary parcellation model yields distributions close to lognormal. (A) A parent region is chosen randomly and divided into two new regions. This process is repeated until a target number of PUs is reached. (B,C) Distributions of PU sizes generated by the model. (B) 100 PUs: σ=1.40; $p_{KS}$=0.93; (C) 1,000 PUs: σ=2.31; $p_{KS}$=0.33. In these and subsequent simulations, immediately after a division, two divided regions are varied in size by 10% randomly, while maintaining the sum of their volumes constant. This was done to produce region sizes different from fractions of 2.

The model presented above generates new PUs by randomly and uniformly choosing a region to be split into two new regions of equal size. Because in this case the region to be split is chosen independently on its size, this mechanism could be called a model of uniform parcellation. We showed that this model leads to a distribution of PU sizes that is close to lognormal (Figure 3), as observed in the experimental data (Figure 2). Can this data be explained by a different model, e.g., when the regions to be split are not chosen with equal probability? It is possible, for example, that the regions to be fragmented are chosen according to their size. The probability of fragmenting a region number $i$ at a given time-step could be described by the power law distribution

$$p_i = v_i^\mu / Z. \qquad (1)$$

Here $v_i$ is the volume of PU number $i$, $Z$ is the normalization coefficient, while $\mu$ is a parameter (exponent) of the model. For $\mu = 0$, the probability of dividing a PU is independent of its size, which describes the model of uniform parcellation we considered before (Figure 3). For positive values of $\mu$, larger regions are more likely to be split, leading to a brain containing PUs of more similar size. For negative $\mu$, smaller PUs are expected to be split more often, leading to a wide distribution of PU sizes. The more general mechanism described by equation (1) will be called here the model of biased parcellation.

To analyze quantitatively the parcellation process for various values of $\mu$ we compute the standard deviation of the logarithm of PU sizes as a function of $\mu$ (Figure 4). The standard deviation defines the width of a PU size distribution. The width of a distribution is dependent on the number of PUs. Indeed, within our model, the number of PUs determines the number of time steps needed to evolve the set of brain structures. Longer evolution leads to a larger diversity in PU sizes, since PU distribution is determined by a diffusion-like process. A simple calculation shows that, for $\mu = 0$, for example, the variance of logarithms of PU sizes depends on time as $\sigma^2(\ln v) \sim \ln t$, where evolutionary time is measured by the number of PUs, $t = N_{PU}$. For any given $\mu$, therefore, the diversity of PU sizes can depend on the number of PUs. The three panels in Figure 4 present the dependence of $\sigma(\ln v)$ on the exponent $\mu$ for the number of PUs $N_{PU}$ matching the three species for which the data is available (Figure 2). By obtaining the range of $\mu$ consistent with the $\sigma(\ln v)$ observed in each of the species, one can infer the exponent guiding brain parcellation in these animals. The shaded region in Figure represents a 90% confidence interval for the values of $\mu$, obtained from a sample of 100 simulations for each data set. The 90% confidence intervals for $\mu$ overlap in the three species within the interval $0.11 \leq \mu \leq 0.22$. This suggests that the evolution of brain parcellation may have followed a similar rule in the three organisms, with the selection of next region for fractioning slightly biased toward larger PUs ($\mu > 0$). The model of uniform parcellation described above (Figure 3) is therefore not so far from reality.

The evolutionary parcellation model described here has implications for the connectivity strengths between different brain regions. Such connectivity is defined through the total number of wires running between a pair of regions and is conventionally called the macroconnectome, to distinguish it from the connectivity with a single-neuron precision, i.e. the microconnectome. In the simplest parcellation model of the macroconnectome, brain evolution starts from the tabula rasa state in which every two neurons have a finite probability $f$ to be connected. The total number of connections is therefore equal to $C = fN^2/2$, where $N$ is the total number of neurons and $N^2/2$ is the total number of pairs. After the parcellation process has run its course, the number of connections between any two areas with volumes $v_i$ and $v_j$ is expected to be $C_{ij} = v_i f \rho^2 v_j / 2$. Here $\rho$ is the number density of neurons, assumed to be the same for all areas (*12*). Within the parcellation model, the macroconnectome defined by matrix $C_{ij}$ is therefore expected to have the outer product form, i.e. $C_{ij} \propto v_i v_j$. This simple theory generates an experimentally testable prediction. In particular, it predicts that the strength of connections for one of the brain regions, number $i$, for example, with an array of other regions, numbered by index $j$, should be proportional to the size of these regions $v_j$, which is a consequence of the outer product form of the connection matrix $C_{ij} \propto v_i v_j$.

To test the outer product form of the macroconnectome, we analyzed the dependence of the number of connections between three macaque visual areas, V1, V2, and V4, and an array of 91 cortical areas. The data was available from a recent study of the distribution of the number of connections (*13*). This data contained the number of neurons projecting to a given target area obtained using retrograde tracing of connections. To eliminate uncertainties associated with the dimensions of tracer injection, the number of connections was normalized to one. Thus the data is represented by a normalized connection matrix $\tilde{C}_{ij} = C_{ij} / \sum_k C_{ik}$, called the fraction of labeled neurons (FLN). It is easy to see that $\sum_k \tilde{C}_{ik} = 1$. If the original connection matrix $C_{ij}$ is given by the outer product form $C_{ij} \sim v_i v_j$, as our model suggests, the FLN matrix is proportional to the volume of the source, i.e. $\tilde{C}_{ij} \propto v_j$. We therefore plotted the logarithm of connection strength $\tilde{C}_{ij}$ as a function of the logarithm of source volume $v_j$ expecting to observe a positive correlation. Indeed, we find that for all three target areas, V1, V2, and V4, the logarithm of normalized non-zero connection strength is correlated with the logarithm of source area volume (Figure 5A). For the area V1, for example, the correlation is close to R=0.78, suggesting that R²=62% of the connectivity data is explained by the source area size. For other areas the correlation is weaker, however, and data for these areas appears to be more variable due to fewer injections (the number of injections yielding connectivity was 5, 4, and 3 for V1, V2, and V4 in Ref. (*13*), producing $R^2 =$ 0.62, 0.54, and 0.25 respectively).

Although area-to-area connectivity appears to be close to the outer product form, the slope of dependences in Figure 5A is not consistent with the simple parcellation model presented above. Indeed, although linear fits to dependences in Figure 5A account for large amounts of data (R²), these linear fits result in a power law relationship between the area-to-area connection strength and source volume, i.e. $\tilde{C}_{ij} \propto v_j^\eta$, with $\eta = 2.67$, 2.85, and 1.79, respectively, for V1, V2, and V4. The exponent obtained from fitting all data for V1, V2, and V4 is $\eta = 2.33$ (R=0.63). In contrast, the simple parcellation model for connectivity presented above yields $\eta = 1$ ($\tilde{C}_{ij} \propto v_j$). Thus, although the simple model accurately predicts the outer product form of area-to-area connectivity and the power-law dependence of connection strength on PU volume ($\tilde{C}_{ij} \propto v_j$), the exponent in the law is not captured exactly; the observed value of the exponent is larger. This is expected, however, since the macroconnectome can be modified to be better suited for the particular computations relevant for an organism's survival after the areas are split. Nonetheless, a substantial amount of data on non-zero macroconnectivity (as much as 62% for area V1) can be explained by the source area size through the power law described above.

To gain insight into the basis of the power law relationship between the number of area-to-area connections and the source area size, we propose a model that is based on a Hebbian learning rule. It was recently proposed that the microconnectome (connections within a cortical column) is determined by the scale-free multiplicative Hebbian learning rules of the form (*14*)

$$dC_{ij}/dt = \varepsilon_1 f_i^\alpha C_{ij}^\beta f_j^\gamma - \varepsilon_2 C_{ij}. \qquad (2)$$

Here, $f_i$ is the activity level in area number $i$, the first term describes the correlations in neural activity, the second term in the r.h.s. is the connection decay, $\varepsilon_1$ and $\varepsilon_2$ are constant parameters, while $\alpha$, $\beta$, and $\gamma$ are exponents. By assuming that connection strengths have already reached equilibrium for a given activity configuration ($dC_{ij}/dt = 0$) and that the overall activity levels are proportional to the area size ($f_i \propto v_i$), we obtain $C_{ij} \propto v_i^{\alpha/(1-\beta)} v_j^{\gamma/(1-\beta)}$, i.e. the outer product power-law form of connectivity with $\eta = \gamma/(1-\beta)$ and $\kappa \equiv \alpha/(1-\beta)$. To match our observations for area V1, we have to assume that $\gamma/(1-\beta) = \eta \approx 2.67$. Thus, the same Hebbian learning rule postulated in (*14*) can be used to derive both micro- (within area) and macroconnectivity (area-to-area).

The outer product form can also be tested for the outgoing connections, i.e. $C_{ij} \sim v_i^\kappa$. Recent studies make available outgoing connectivity strengths for 295 target mouse brain regions (*15*). This data suggests that the outer product form of connectivity describes $R^2$=30-50% of connection data with an exponent $\kappa \approx 1.3$ (Figure 5B,C). Data in Figure 5B,C are presented for four regions with $\geq 3$ injections. Although Figure 5 includes only non-zero connections, it is notable that the Hebbian learning rule (2) includes zero connections ($C_{ij} = 0$) as a solution, thus capturing both vanishing and non-vanishing connections.

Here, we have studied an evolutionary model for the emergence of a distribution of brain regions (PUs) according to their size. We aimed at describing the ensemble of brain regions statistically, without addressing the functional significance of individual PUs. We assumed that brain regions emerge through a sequential process of fragmentation and specialization. Kolmogorov's fragmentation model (*6*) assumes that the splitting process for each fragment occurs independently of other pieces. This model is appropriate to describe explosive shell fragmentation or rock grinding (*3-5*). For such a process, under conditions of stationary parameters, the total number of pieces grows exponentially. Since there is no evidence for the explosion in the number of brain regions that occurred recently, we were motivated to find a different model that would describe a more gradual proliferation of PUs, one at a time.

We find that, similar to Kolmogorov's model, the distribution of PU sizes in our model is close to lognormal. The variance of the logarithm in our model depends on the scaling of the splitting probability with volume [equation (1)]. We find that in all three species, mouse, macaque, and human, brain/cortex parcellation is consistent with a scaling exponent within the range $0.11 \leq \mu \leq 0.22$, suggesting that common evolutionary mechanisms may have shaped the brains of these animals. It is conceivable that each PU undergoes multiplicative variations in size between fragmentations, which should slightly increase the values of $\mu$. We find therefore that the probability of PU fragmentation is dependent on PU size [equation (1)], by contrast with the Kolmogorov theory.

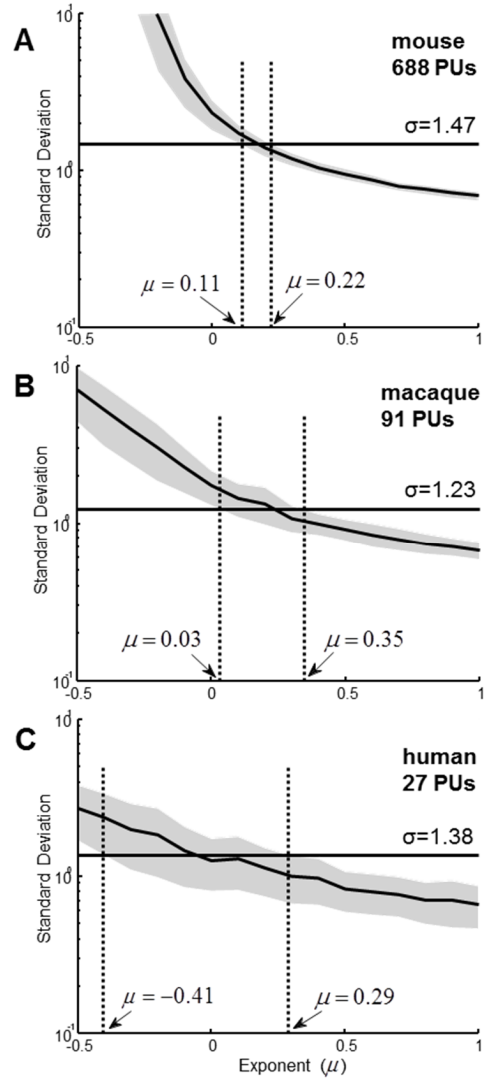

**Figure 4.** The standard deviation of the logarithm of PU size as a function of exponent $\mu$. The shaded region represents a 90% confidence interval for $\mu$, which was obtained from a sample of 100 simulations for each data set. The horizontal line describes the observed value. Ranges consistent with the model are indicated by the dotted lines. (A) Mouse brain. $0.11 \leq \mu \leq 0.22$. (B) Macaque cortex: $0.03 \leq \mu \leq 0.35$. (C) Human cortex: $-0.41 \leq \mu \leq 0.29$.

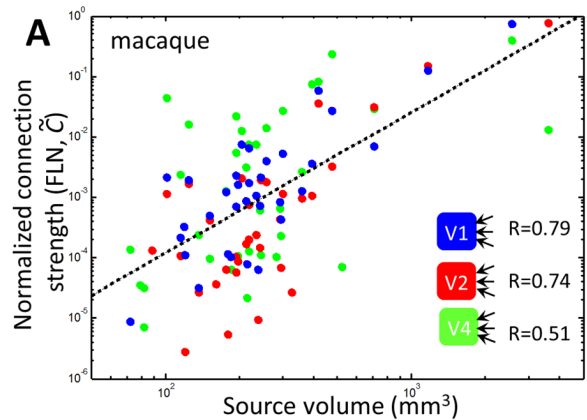

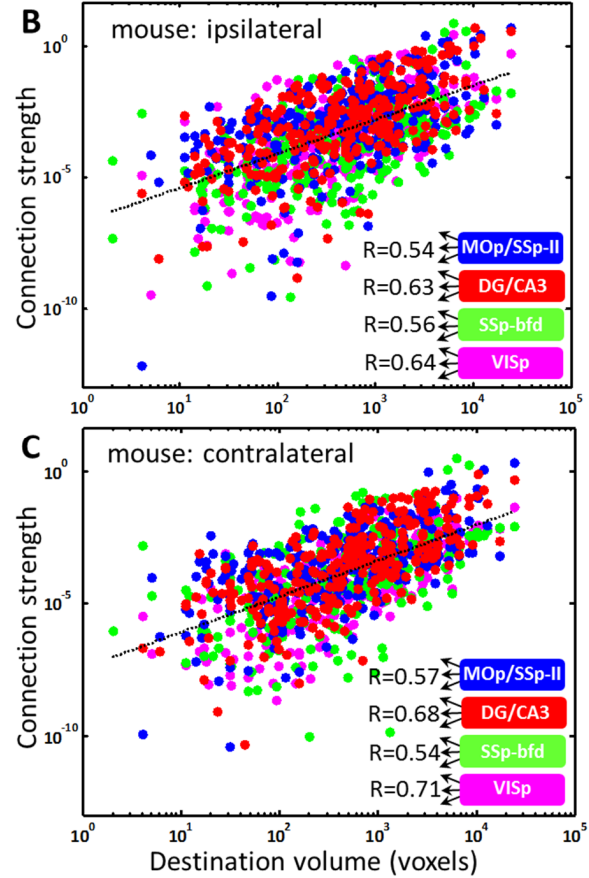

**Figure 5.** PU volumes contain information about connection strengths. The dependences of connection strength on PU size. (A) Non-vanishing incoming connection strengths for three target visual areas in the macaque brain and various other areas as sources. The linear fit (dotted line) has a slope of $\eta = 2.33$. Individual slopes for the three areas V1, V2, and V4 are $\eta = 2.67$, 2.85, and 1.79. The scaling law $\tilde{C}_{ij} \propto v_j^\eta$ can explain the fractions $R^2 = 0.62$, 0.54, and 0.25 data variance for these areas. (B and C) Non-vanishing outgoing connections in the mouse brain for four source regions are described by a similar scaling relationship $C_{ij} \propto v_i^\kappa$, where the exponent $\kappa = 1.30$ and 1.35 for ipsi- and contralateral connections is obtained from the linear fits (dotted lines).

We analyzed the dependence of area-to-area (macro) connectivity in the macaque cortex and its relation to PU sizes that was suggested by our parcellation model. We found that the incoming connectivity is well described by the outer product power-law form with similar scaling relationships for three visual areas: V1, V2, and V4. Remarkably, we find that for area V1, 62% of information about its non-zero incoming connection strengths to other areas is contained in the simple scaling relation $\tilde{C}_{ij} \propto v_j^\eta$, with the scaling exponent $\eta > 1$. Macro connection strength is therefore mostly determined by the sizes of connected regions. The lognormal distribution of connection strengths (*13*) is therefore a consequence of the lognormal distribution of PU sizes, reported here. The remaining 38% of V1 connectivity may be attributed to non-outer product features, such as connection locality (*13, 16*), as well as residual noise in data. We find power law $\tilde{C}_{ij} \propto v_j^\eta$ fit quality is dependent on the number of injections into targets, suggesting that more reliable data could improve fit quality. We proposed a Hebbian learning paradigm that could explain the outer product form of macroconnectivity. We thus proposed a universal evolutionary law that could both guide the formation of brain regions and contribute substantially to their connectivity. This law generalizes across several mammalian species.

**References**

1. J. Kaas, H., *Evolutionary Neuroscience.* (Academic Press, 2009).
2. G. F. Striedter, *Principles of Brain Evolution.* (Sinauer Associates, 2004).
3. B. Epstein, in *Journal of the Franklin Institute.* (1947), pp. 471-477.
4. D. E. Grady, M. E. Kipp, in *Journal of Applied Physics.* (1985), pp. 1210-1222.
5. S. Redner, in *Disorder and Fracture.* (Plenum Press, New York, 1989), pp. 31-50.
6. A. N. Kolmogorov, On the lognormal law of distribution of particle sizes in fragmentation *Proceedings of the Academy of Sciences of USSR* **31**, 99 (1941).
7. P. Osten, Y. Kim. (Cold Spring Harbor, NY, USA, 2013).
8. R. Bakker, G. Bezgin, R. Kötter, in *Scalable Brain Atlas.* (Donders Institute, Radboud University and Medical Center Nijmegen, 2014).
9. J. K. Mai, G. Paxinos, V. Thomas, *The Atlas of the Human Brain.* (Academic Press, 2007).
10. K. H. Rosen, J. G. Michaels, *Handbook of discrete and combinatorial mathematics.* (CRC Press, Boca Raton, 2000), pp. 1232 p.
11. F. J. Massey, The Kolmogorov-Smirnov Test for Goodness of Fit. *J Am Stat Assoc* **46**, 68 (1951).
12. C. N. Carlo, C. F. Stevens, Structural uniformity of neocortex, revisited. *P Natl Acad Sci USA* **110**, 1488 (Jan 22, 2013).
13. N. T. Markov *et al.*, Weight Consistency Specifies Regularities of Macaque Cortical Networks. *Cereb Cortex* **21**, 1254 (Jun, 2011).
14. A. A. Koulakov, T. Hromadka, A. M. Zador, Correlated Connectivity and the Distribution of Firing Rates in the Neocortex. *J Neurosci* **29**, 3685 (Mar 25, 2009).
15. S. W. Oh *et al.*, A mesoscale connectome of the mouse brain. *Nature* **508**, 207 (Apr 10, 2014).
16. V. A. Klyachko, C. F. Stevens, Connectivity optimization and the positioning of cortical areas. *P Natl Acad Sci USA* **100**, 7937 (Jun 24, 2003).